\documentclass[12pt,preprint]{aastex}
\usepackage{amssymb}

\shorttitle{De Buizer, Pi\~{n}a, \& Telesco}
\shortauthors{Mid-IR Observations of G339.88-1.26}

\begin{document}

\title{High-Resolution Mid-Infrared Imaging of G339.88-1.26}
\author{James M. De Buizer\altaffilmark{1,2}, Andrew J. Walsh\altaffilmark{3}, 
Robert K. Pi\~{n}a\altaffilmark{2,4}, Chris J. Phillips\altaffilmark{5},
and Charles M. Telesco\altaffilmark{2,4}}

\altaffiltext{1}{Cerro Tololo Inter-American Observatory, National Optical Astronomy
Observatory, Casilla 603, La Serena, Chile.
CTIO is operated by AURA, Inc.\ under contract to the National Science
Foundation.} \altaffiltext{2}{Visiting Astronomer, W.M. Keck Observatory.} 
\altaffiltext{3}{Max-Planck-Institut fuer Radioastronomie, Auf dem Huegel
69, D-53121, Bonn, Germany.} 
\altaffiltext{4}{Department of Astronomy, 211 Space Sciences Research Building, University of
Florida, Gainesville, FL 32601.} 
\altaffiltext{5}{Joint Institute for VLBI in
Europe, Radiostrerrenwacht Dwingeloo, Postbus 2, 7990 AA Dwingeloo, The
Netherlands.}

\begin{abstract}
G339.88-1.26 is considered to be a good candidate for a massive star with a
circumstellar disk. This has been supported by the observations of linearly
distributed methanol maser spots believed to delineate this disk, and
mid-infrared observations that have discovered a source at this location
that is elongated at the same position angle as the methanol maser
distribution. We used the mid-infrared imager/spectrometer OSCIR at Keck to
make high-resolution images of G339.88-1.26. We resolve the mid-infrared
emission into 3 sources within 1.5 arcsec of the location of the masers. We
determine that the methanol masers are most likely not located in a
circumstellar disk. Furthermore we find that the observed radio continuum
emission most likely comes from two sources in close proximity to each
other. One source is an unobscured massive star with an extended HII region
that is responsible for the peak in the radio continuum emission. A second
source is embedded and centered on the elongation in the radio continuum
emission that is believed to be tracing an outflow in this region.
\end{abstract}

\keywords{infrared: ISM --- ISM:individual(G339.88-1.26) --- stars:
early-type --- stars: formation --- HII regions --- masers --- ISM:jets and
outflows}

\section{Introduction}

It is generally accepted that stars form via accretion processes, and they
achieve their final masses predominantly through accretion of star-building
material from a circumstellar disk (for a review, see Boss 2000). It is from
these accretion disks that planetary systems are thought to eventually form.
Observations of accretion disks and accreting protostars have supported this
hypothesis in the case of low and solar mass stars \citep{Boss00}. However
there has been no such observational support of the accretion disk scenario
in the high mass portion of the stellar mass spectrum. Some authors suspect
that high mass stars are created through a different process than low mass
stars, like low-mass protostellar mergers, for instance (for a review, see
Garay and Lizano 1999).

On the other hand, circumstantial evidence of circumstellar disks around
massive stars does exist in the form of radio wavelength emission from
outflows and masers. Outflow is an important indirect indicator of the
presence of a circumstellar disk. In the accretion paradigm, when the star
is gaining mass via accretion from a disk it is also accompanied by a
bipolar outflow that is collimated by the disk. The presence of a disk may
also be ascertained from observations of molecular masers. Maser emission
from various molecular species has been observed for several decades and is
a well-known indicator of recent massive star formation. These masers occur
in spatially localized regions or ``spots'' and serve as probes of the
small-scale structure, dynamics, and physical conditions of the environments
near forming stars. Radio observations by \citet{Norris98} and \citet{Phil98}
have shown that methanol (CH$_{3}$OH) maser spots are frequently distributed
in linear patterns, with projected dimensions typically spanning 2500 AU.
Furthermore, in many cases the velocities of the individual maser spots show
a linear trend across the source, indicative of the masers tracing a
rotating structure. It has thus been plausibly argued that in these cases
the methanol maser spots occur in, and directly delineate, rotating
circumstellar disks. However, other authors \citep{SD94, wbhr98} suggest
such linear arrangements of masers, and even their systemic velocities, may
be explained by shock models.

G339.88-1.26 is one such site of linear methanol masers. It was observed to
have ionized material elongated at a position angle perpendicular to the
maser spot distribution angle by \citet{enm96}, who claim the masers are
located within a disk. This hypothesis seems to have been reaffirmed by the
first mid-infrared observations of this source by \citet{sk98}, which showed
a 10 $\micron$ source elongated at the same position angle as the methanol
masers. It was suggested that this was thermal emission from the dust in the
circumstellar disk in which the methanol masers existed. However,
independent mid-infrared observations in a survey by \citet{DPT00} showed
that the source had more structure at 18 $\micron$, resolving what appeared
to be a companion source lying at the same position angle as the source
elongation.

To date, G339.88-1.26 is considered to be one of the best candidates for
being a massive star with a circumstellar accretion disk, and to be good
evidence that methanol masers exist in circumstellar disks. In this paper we
present new high resolution images of G339.88-1.26 taken from the W.M. Keck
telescope. The increased angular resolution reveals details not seen before
in the mid-infrared, and uncovers a region of a different nature than
previously thought.

\section{Observations and Data Reduction}

Observations were obtained at Keck II between 11 and 13 hours UT on 30 April
1999 and again between 12 and 13 UT on 1 May. The University of Florida
mid-infrared camera/spectrometer OSCIR was used for all observations. OSCIR
employs a Rockwell 128x128 pixel Si:As BIB (blocked impurity band) detector,
with a 0\farcs0616 pixel$^{-1}$ scale at Keck II. The total field of view of
the array is 8$\arcsec\times $8$\arcsec$. Images were taken through two
filters, $N$ ($\lambda _{o}$=10.46 $\micron$, $\Delta \lambda $=5.1 $\micron$%
) and $IHW18$ ($\lambda _{o}$=18.06 $\micron$, $\Delta \lambda $=1.7 $\micron
$). Images were centered on a methanol maser reference feature at
R.A.(J2000)=16$^{h}$52$^{m}$04\fs66, Decl.(J2000)=-46$\arcdeg$08$\arcmin$34%
\farcs2. Keck images presented in this paper are composites made up of
several images (2 for $N$, 4 for $IHW18$) with on-source exposure times of
120 seconds each, which were then registered and stacked to improve the
signal-to-noise. Background subtraction was achieved during observations via
the standard chop-nod technique. Flux density calibration was achieved by
observing at a similar airmass the mid-infrared standard star $\eta $ Sgr,
for which the flux densities were taken to be 188 Jy at $N$ and 66 Jy at $%
IHW18$. Point-spread function (PSF) stars, were also imaged near the
position of G339.88-1.26, yielding a measured full width at half maximum
(FWHM) of 0\farcs33 at $N$ and 0\farcs41 at $IHW18$. Because of time
constraints, there was no attempt to perform accurate astrometry for the
observations.

Using the 10 and 18 $\micron$ data, we constructed dust color temperature
(T) and emission optical depth ($\tau $) maps for each source. To do so, the
10 $\micron$ source image was convolved with a 18 $\micron$ point-spread
function (PSF) and the 18 $\micron$ source image with a 10 $\micron$ PSF so
that both images are at exactly the same spatial resolution. Next, we
employed two techniques to find the best spatial registration of the two
images. First, we used an automated registration algorithm based on
minimizing the sum of the squared residuals of the image difference as a
function of the relative offsets. This algorithm generates a ``chi-squared''
surface at integral x and y pixel offsets. The chi-squared surface may then
be interpolated to determine the location of the minimum to fractional pixel
values. The second method was to simply overlay the two images and acquire
the best perceived registration visually. This alignment was a relatively
easy task due to the source brightnesses and similar structure seen at both
wavelengths. Offsets found from these two techniques were in agreement to
within 2 pixels. We therefore estimate the alignment between the 10 and 18 $%
\micron$ images to be accurate to better than 0\farcs1 (2 pixels).

Once the image set was spatially registered, the 10 and 18 $\micron$ flux
densities for each pixel were used to iteratively solve for emission optical
depth ($\tau $) at 10 $\micron$ and dust color temperature. For these
calculations we used the relationship $\tau _{10\micron}=\tau _{18\micron%
}/1.69$ (from the extinction law of Mathis 1990). Tests showed that shifts
in alignment between the two convolved images within the above quoted 2
pixel error can change the peak temperature and optical depth values by $\pm
7\%$. Furthermore, these shifts can create slight changes ($<0\farcs2$) in
peak locations, however the overall trends in the temperature and optical
depth distributions are preserved.

\section{Results and Discussion}

Presented in Figure 1a and 1b are the 10 and 18 $\micron$ images,
respectively, from Keck. These images show only the source referred to as
G339.88-1.26:DPT00 1 from \citet{DPT00}. One can see that the simple
elongated source first seen by \citet{sk98} with the European Southern
Observatory 3.6-m\ telescope, is now resolved into three mid-infrared
components. The western source seen in these Keck images, which we will
refer to as G339.88-1.26:DPT00 1C, is the same as that resolved by %
\citet{DPT00} on the Cerro Tololo Inter-American Observatory\ 4-m with
OSCIR. However, the Keck images show that there are two sources to the east
of 1C\footnote{%
We will use a shortened form of the IAU recommended names, i.e.
G339.88-1.26:DPT00 1C will be referred to as 1C.}. The easternmost source is
the brightest at 18 $\micron$ (G339.88-1.26:DPT00 1A), but the flux density
peak at 10 $\micron$ is the source between 1A and 1C (G339.88-1.26:DPT00
1B). Finding integrated flux densities for these sources is difficult
because they each display extended morphologies, and are embedded in a
region of more extended emission. The integrated flux densities tabulated in
Table 1 are found by using a two-dimensional Gaussian model to subtract off
one source in the pair and then performing aperture photometry on the
second. Given this uncertainty, combined with atmospheric variability and
uncertainty in the standard star flux density, we quote the flux density
values for these sources in Table 1 with a 10\% photometric error.

The eastern source 1A is elongated at a position angle of $\thicksim $315$%
\arcdeg$ (more clearly seen in the 18 $\micron$ image). The central source
1B may be extended, as seen in the 10 $\micron$ image (Figure 1a), however
the surface brightness distribution in this source may be confused by the
close proximity of 1A. Source 1C has a diffuse appearance, and the emission
appears as if dust has been swept up to the northwest in a `tail', an effect
seen predominantly in the 18 $\micron$ image.

\subsection{The Nature of the Sources}

The dust color temperature and optical depth maps are shown in Figure 1c and
1d, respectively. The temperature map shows that the temperature peak ($%
\thicksim $145 K) is 0\farcs4 due west of the mid-infrared peak of source
1B, and that the temperature drops off uniformly to the west and east. Clear
temperature peaks, such as this one, indicate the positions of the stellar
heating sources. The optical depth map shows that emitting material here is
optically thin at 10 $\micron$, and that the optical depth is higher for 1A
and 1C than for 1B.

Qualitatively, the mid-infrared surface brightness distribution of the
sources appear similar to the observations of the circumstellar disk of HR
4796A \citep{J98, K98}. Analogous to the mid-infrared observations of
HR4796, G339.88-1.26 has three aligned mid-infrared sources with a
temperature peak near the central mid-infrared source, delineating the
stellar source. Likewise, the eastern and western sources in G339.88-1.26
can be thought of as being analogous to the lobes of a disk. However, the
disk scenario seems unlikely for several reasons. First, at 18 $\micron$,
sources 1A and 1C are significantly different in brightness, and it would be
difficult in disk models to explain such a disparity in the flux densities
on opposite sides of a circumstellar disk. Furthermore, the wispy, diffuse
appearance of 1C is an unlikely morphology for a lobe of a disk, and it
would be difficult to explain this tail of emission rising northward.
Finally, the temperature peak is offset from the mid-infrared peak of 1B. As
discussed in the section on observations and data reduction, shifts in the
temperature peak can be caused by improper registration of the 10 and 18 $%
\micron$ convolved images. However, tests showed that there was no alignment
of the two images within our estimated error that could produce a
temperature peak at the location of the mid-infrared peak of source 1B, and
the predominant change in temperature peak position ($<0\farcs2$) in these
trials was in the north-south direction. Therefore, we are confident that
the offset between the temperature peak and mid-infrared peak of 1B is real.
Consequently, it seems unlikely that 1B is the circumstellar dust
surrounding the stellar source present at this temperature peak.

So what is responsible for the temperature peak? There is a radio continuum
source at the location of G339.88-1.26:DPT00 1, seen at 8.6 GHz by %
\citet{enm96} and \citet{wbhr98}. \citet{wbhr99} discovered a source at H
(1.65 $\micron$) and K (2.2 $\micron$), that is coincident with the location
of the radio continuum peak. Furthermore, this source seen in the near
infrared can also be seen in the Digitized Sky Survey at visible
wavelengths. An important question to answer is: does the optical/near
infrared source and the radio continuum source have the same origin? It is
certainly plausible from the fact that the optical, near infrared and radio
sources are all coincident within their respective positional errors.
However, the optical/near infrared source may be photospheric emission from
a foreground object, with the radio source coming from a second deeply
embedded object. We can rule out a further possibility that there is only
one deeply embedded object, because we can see the source optically.

We can use our mid-infrared observations to help clarify the situation. We
see a temperature peak in Figure 1c indicating the presence of a source
hotter than those sources seen in the mid-infrared. The
optical/near-infrared emission therefore most likely originates from the
location of the dust temperature peak. If this optical/near infrared source
is a foreground source, we surmise that it must be close enough to the dusty
material we see in the mid-infrared to directly heat it. We do not see any
evidence for a peak in either the 10 or 18 $\micron$ dust emission maps
(Figure 1a and 1b) at the position of the temperature peak, something we
would assume from an embedded UCHII region. Therefore we have no evidence
that there is a second embedded source at the same position as the
optical/near infrared source. Thus, we conclude that the most plausible
scenario is that the optical, near infrared and radio continuum emission all
come from a star located just in front of the mid-infrared sources.

At first glance, it appears contradictory that we find compact radio
continuum emission around a star that can be seen in the optical. However,
the radio continuum observations by Ellingsen et al. (1996; shown in Figure
2) were made using the 6A configuration of the Australia Telescope Compact
Array (ATCA), which is insensitive to large structures. Further
observations, using a more compact configuration of the ATCA (Ellingsen,
private communication), show resolved emission extending over approximately
30$\arcsec$. We estimate the spectral type of the star to be B2.5, based on
the integrated radio flux of 14 mJy and a distance of 3.1 kpc. The Str\"{o}%
mgren radius of a B2.5 star is expected to be 0.3 pc, assuming an electron
density of 10$^{-4}$ cm$^{-3}$. At a distance of 3.1 kpc, this is equivalent
to a diameter of 40$\arcsec$. Thus the size of the ionized region agrees
well with that expected from an unembedded HII region around a B2.5 star.
Since the source responsible for this radio peak is also seen at wavelengths
as short as visible, this star could not be significantly embedded in
circumstellar material. In fact, the mid-infrared emission optical depth
decreases to the lowest value at this location.

It therefore appears that there are 4 different sources in this small
region: 1A, 1B, 1C and the optical/near-infrared/radio continuum source,
which is unobscured and located at the temperature peak. Though we have no
information on the three dimensional distribution of the sources in this
region, the close 2 dimensional proximity of the dusty source 1B to this
unobscured stellar source indicate that it is perhaps slightly foreground to
1B. A possible scenario is that the optical/near infrared/radio source
formed from the same molecular cloud containing the other sources, but is
located on the edge closest to us.

In Figure 2a, we show radio continuum data from \citet{enm96}. From this map
it is easy to see that the continuum is elongated and that there are also
other knots in the radio map extending far beyond the compact center. Line
cuts in the x pixel direction were made through the map of the radio source
to find the peak flux density along the source as a function of pixel
position y. A linear regression line was fit to these points, yielding a
position angle for the elongation. The continuum around the radio peak was
found to be elongated at a position angle of 46$\arcdeg$, and it appears
that the separate knots of radio continuum extend out in both directions
along this axis. Given the collimated nature of the radio continuum, we
suggest that it is likely to be tracing an ionized outflow. In Figure 2b we
show the 10 $\micron$ contours, overlaid with the radio continuum contours
centered on the dust temperature peak location. However, it can be seen in
this figure that the collimated outflow axis does not bisect the radio peak,
but instead bisects the mid-infrared source 1B. Thus, it seems there are two
components contributing to the radio continuum emission: an ionized outflow
from 1B and a second component aligned with the temperature peak. This
strengthens our astrometric argument that the radio continuum peak is
located at the dust temperature peak. This is the only registration of the
radio continuum peak that yields a source on the outflow axis.

We suggest that source 1B is an embedded protostar (since we see no
temperature peak at its location) with an ionized outflow. Interestingly,
the 10 $\micron$ image of 1B shows that it may be elongated at a position
angle roughly perpendicular to the axis of outflow. This elongation may be
due to a disk, however we caution that it may also be due to slight
differences in the positions of the individual frames that were stacked to
make the final image, and/or confusion of extended emission from source 1A.
The possible elongation of 1B cannot be confirmed at 18 $\micron$ because of
the confusion with source 1A. In Figure 2b, it can be seen that there are
knots of radio continuum emission located equidistant from the location of
1B (marked by crosses), which may have been bipolar mass loss events in the
past. This again strengthens our astrometric argument that 1B is the center
of the outflow and that the radio continuum peak is associated with the dust
temperature peak (and thus, the optical/near infrared source). The presence
of ionized components of the outflow in such close proximity to the location
of 1B implies that the protostar here is still actively outflowing. Assuming
the outflowing material is moving at the speed of a standard high velocity
(SHV) outflow of $\thicksim $20 km sec$^{-1}$ \citep{B96}, the age of the
closest outflow event to 1B would be approximately 1300 yr, given a distance
of 3.1 kpc \citep{whrb97}. However, ionized radio jets from massive stars
have been observed with velocities a magnitude or more than this value %
\citep{R96}, so these outflow events may have been much more recent. The
central elongation of the radio continuum is more prominent to the northeast
than to the southwest, perhaps due to inclination effects, or perhaps due to
anisotropy in the parent molecular cloud, however there are also some bright
knots of ionized material seen along the outflow axis to the southwest.

In Figure 1d, we see that the eastern optical depth peak follows the 18 $%
\micron$ flux density contours of source 1A quite well. Since this source
has a low emission optical depth at 10 $\micron$, and since there is no
temperature peak at the location of 1A, it is likely that this source is
bright in the mid-infrared simply because we are seeing through a large
column of mid-infrared emitting material. This knot of material may be
heated externally by either 1B or the optical/near infrared stellar source.

This same scenario seems likely for 1C, as well. However, 1C is
morphologically distinct in that it has a wispy, coma-like appearance. One
cannot help but draw morphological similarities to mid-infrared observations
of AFGL 2591 \citep{mjf00}. This site contains two sources, AFGL 2591A and
AFGL 2951B, with the latter displaying a morphology similar to
G339.88-1.26:DPT00 1C. One scenario given for the morphology of source AFGL
2951B is that it is a protostar whose envelope may be affected by an
observed outflow from AFGL 2951A, creating the `coma'-like shape. In the
same manner, it is a possibility that a less-collimated molecular component
of the outflow coming from G339.88-1.26:DPT00 1B along the axis of the
ionized component may be responsible for the morphology of
G339.88-1.26:DPT00 1C.

\subsection{The Methanol Masers}

The relative astrometry between the radio continuum and methanol masers
presented in Figure 2 is accurate to 0\farcs2 \citep{P01}. The methanol
masers plotted were observed using the Australian VLBI network \citep{P01}.
These observations are more sensitive to the weaker masers than the
observations of \citet{N93}, and show 49 masers in the distribution. Because
of the higher spatial resolution, the relative positions of the maser
components are much more accurate and do not suffer from spatial or spectral
blending. The majority of the methanol masers are strung along in a
distribution that runs from the east and curves up to the north. There is
also a western group of methanol masers that are linearly distributed in a
east-west fashion.

What is it that the methanol masers are delineating? The string of masers
(not including the western group) may be tracing a disk (as suggested by
Norris et al. 1998) around the star at the center of 1B, but the masers in
this string lie on average 0\farcs6 from the mid-infrared peak, given our
astrometry. A necessary condition for masers to be in an edge-on disk is
that they are coincident with the mid-infrared peak denoting the center of a
circumstellar dust distribution. If our astrometry is correct, it would be
unlikely that the methanol masers are located in an edge-on disk. This
string of methanol masers tends to follow the radio continuum contours well.
However, they just as likely could be said to appear to trace the outer
mid-infrared contours of 1B. Therefore a more likely explanation is that
these methanol masers are tracing a shock or density enhancement, perhaps in
the material around 1B.

The western group of methanol masers form a line that is coincident with the
brightest part of source 1C. It is not clear how they may be excited.
However, it is possible that they are shock induced as well by the outflow
from 1B into the material of 1C.

\section{Conclusions}

We have resolved what was once thought to be a single circumstellar disk
into three mid-infrared sources near the location of the methanol masers of
G339.88-1.26. Furthermore, a re-inspection of the radio continuum
observations of \citet{enm96} reveals a well-collimated distribution of
radio sources, which we claim is due to an ionized outflow. This outflow
axis does not bisect the radio continuum distribution peak, and it therefore
seems likely that the source responsible for the ionized outflow is not the
same source responsible for the radio continuum peak.

The true relative alignment between the radio and mid-infrared sources is
unknown, and mid-infrared observations with more accurate astrometry are
needed. However, using all of the available observations of G339.88-1.26 we
have found the most probable astrometric scenario. In this scenario it
appears that this small region not only contains the three mid-infrared
sources, but also includes a non-mid-infrared emitting stellar source seen
in the optical and near infrared. It is coincident with the location of the
radio continuum peak, and we therefore argue that this stellar source has an
associated HII region and is probably a massive star of spectral type B2.5.
We also find that the axis of the elongated radio continuum bisects the
mid-infrared source 1B, which appears to lie at the center of, and be
responsible for, the ionized outflow. Given our astrometry, the methanol
masers appear offset from the peaks of the mid-infrared sources, and
therefore it is most probable that the methanol masers are not tracing an
edge-on disk in this case, as claimed by \citet{DPT00}, \citet{D00}, and %
\citet{enm96}. A majority of the methanol masers seem to be associated with
source 1B, though our astrometry places a grouping of methanol masers at the
location of 1C. It is unclear exactly how the methanol masers in this region
are excited, however based on purely morphological arguments for source 1B,
we suggest a more probable supposition is that they lie in the surrounding
dusty envelope.

\acknowledgments The authors would like to thank Simon Ellingsen for the
generous use of his radio data. Some of the data presented herein were
obtained at the W.M. Keck Observatory, which is operated as a scientific
partnership among the California Institute of Technology, the University of
California, and the National Aeronautics and Space Administration. The
Observatory was made possible by the generous support of the W.M. Keck
Foundation.

\begin{deluxetable}{lcccc}
\tabletypesize{\small}
\tablewidth{0pt}
\tablecaption{Mid-Infrared Photometry \label{tbl-1}}
\tablehead{
\colhead{Source} & \colhead{10 $\micron$ Integrated}   & \colhead{10 $\micron$ Peak} & \colhead{18 $\micron$ Integrated} & \colhead{18 $\micron$ Peak} \\
\colhead{Name}   & \colhead{Flux Density}   & \colhead{Flux Density} & \colhead{Flux Density} & \colhead{Flux Density} \\
\colhead{}   & \colhead{(mJy)}   & \colhead{(mJy)} & \colhead{(mJy)} & \colhead{(mJy)}
} 
\startdata
G339.88-126:DPT00 1A &358 ($\pm36$)  &1.0 ($\pm0.1$) &7964 ($\pm796$) &13.0 ($\pm1.3$) \\
G339.88-126:DPT00 1B &202 ($\pm20$) &1.3 ($\pm0.1$) &934 ($\pm93$) &7.8 ($\pm0.8$) \\
G339.88-126:DPT00 1C &99 ($\pm10$) &0.6 ($\pm0.1$) &3464 ($\pm346$) &4.4 ($\pm0.4$) \\
\enddata
\end{deluxetable}

\clearpage

\begin{figure}
\epsscale{0.9}
\plotone{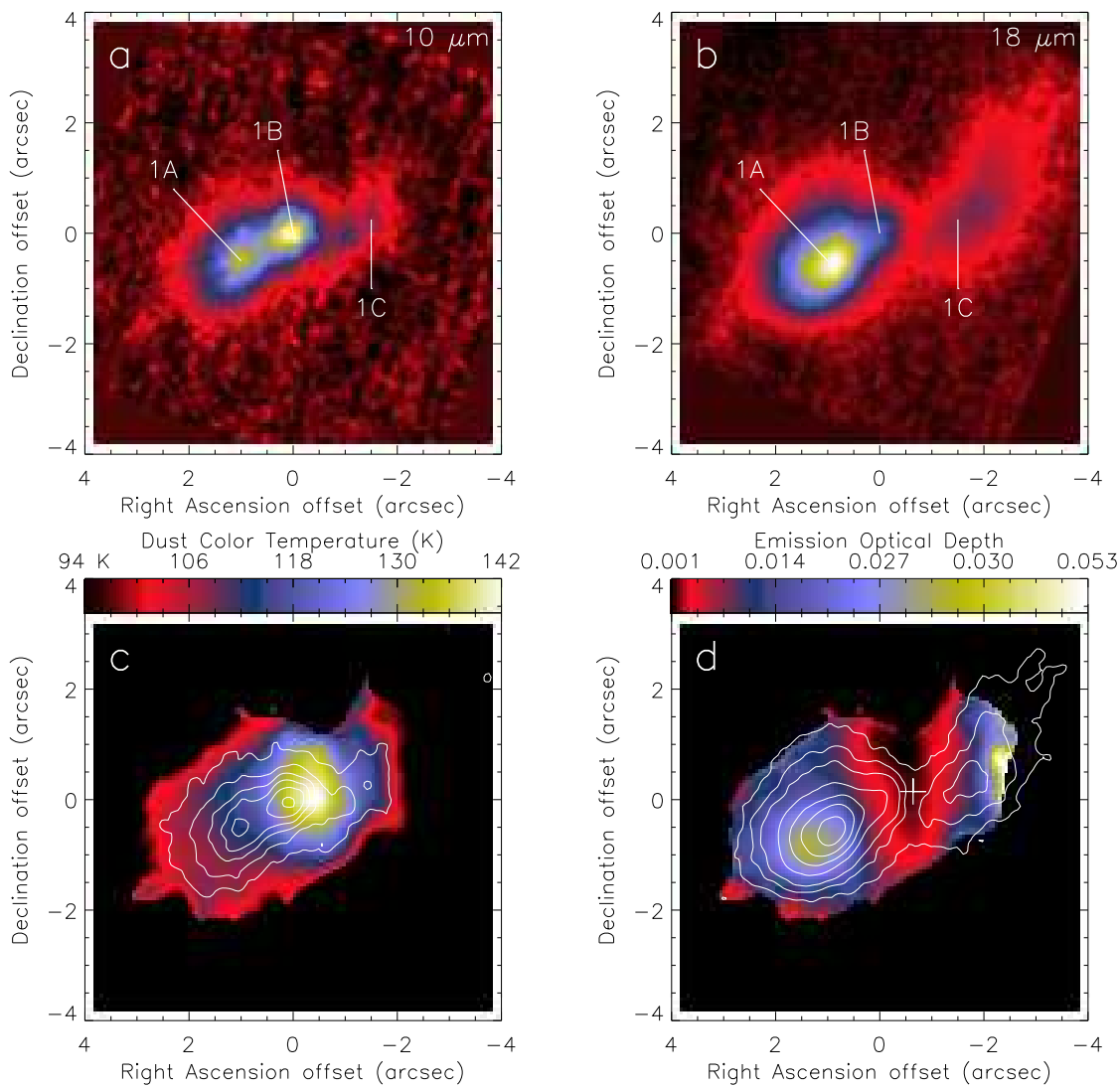}
\caption[f1.eps]{Color OSCIR images from
Keck at 10 $\micron$ (a) and 18 $\micron$ (b), resolve the
source into three mid-infrared components labeled 1A, 1B, and 1C. The dust
color temperature map (c) is overlaid with the 10 $\micron$ contours,
and shows that the temperature peak is 0.4$\arcsec$ west of source 1B. The
emission optical depth map (d) is overlaid with the 18 $\micron$ contours. 
A cross marks the temperature peak location. \label{fig1}}
\end{figure}

\begin{figure}
\epsscale{0.9}
\plotone{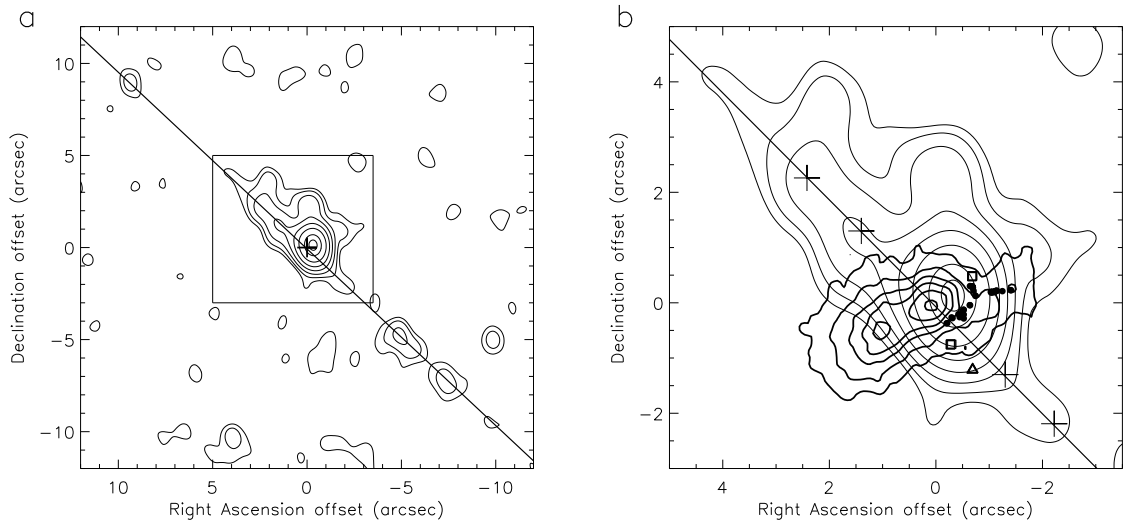}
\figcaption[f2.eps]{Radio continuum emission from G339.88-1.26. (a) This panel shows a wide field 
of view contour map of the 8.59 GHz radio continuum data of Ellingsen et al. (1996). 
The emission around the radio continuum peak is elongated, and there are other knots 
of radio emission lying along the elongation axis (diagonal line). We suggest that this 
radio emission marks an ionized outflow. (b) This panel shows the probable relationship 
between the radio and mid-infrared emission. The 10 $\micron$ contours (thick) are shown 
overlaid with the radio continuum contours (thin, shown in the box in panel a), assuming the 
radio continuum peak is coincident with the optical/near infrared stellar source at the 
location of the color temperature peak. Overplotted are the methanol masers (black 
circles)  from Phillips et al. (2001), the water maser (triangle) from Forster and
Caswell (1989), and the OH masers (squares) from Caswell et al. (1995). 
The outflow axis does not intersect 
the radio continuum peak, but instead the mid-infrared peak of source 1B. The crosses 
mark the positions of two possible bipolar mass loss events that are equidistant from 
the location of 1B.
\label{fig2}}
\end{figure}

\end{document}